\documentclass[twocolumn]{aastex631}

\usepackage{graphicx}
\usepackage[varg]{txfonts}
\usepackage{cases}
\usepackage{natbib}
\usepackage{multirow}
\usepackage{longtable}
\usepackage{graphicx}
\usepackage{booktabs}

\shorttitle{Contraction, expansion, and vertical oscillation of coronal loops}
\shortauthors{Zhang et al.}

\graphicspath{{./}{figures/}}

\begin{document}

\title{First detection of transverse vertical oscillation during the expansion of coronal loops}

\correspondingauthor{Qingmin Zhang}
\email{zhangqm@pmo.ac.cn}

\author[0000-0003-4078-2265]{Qingmin Zhang}
\affiliation{Key Laboratory of Dark Matter and Space Astronomy, Purple Mountain Observatory, CAS, Nanjing 210023, People's Republic of China}

\author[0000-0001-7693-4908]{Chuan Li}
\affiliation{School of Astronomy and Space Science, Nanjing University, Nanjing 210023, People's Republic of China}
\affiliation{Key Laboratory of Modern Astronomy and Astrophysics (Nanjing University), Ministry of Education, Nanjing 210023, People's Republic of China}

\author[0000-0002-4538-9350]{Dong Li}
\affiliation{Key Laboratory of Dark Matter and Space Astronomy, Purple Mountain Observatory, CAS, Nanjing 210023, People's Republic of China}

\author[0000-0002-1190-0173]{Ye Qiu}
\affiliation{School of Astronomy and Space Science, Nanjing University, Nanjing 210023, People's Republic of China}
\affiliation{Key Laboratory of Modern Astronomy and Astrophysics (Nanjing University), Ministry of Education, Nanjing 210023, People's Republic of China}

\author[0000-0003-1979-9863]{Yanjie Zhang}
\affiliation{Key Laboratory of Dark Matter and Space Astronomy, Purple Mountain Observatory, CAS, Nanjing 210023, People's Republic of China}
\affiliation{School of Astronomy and Space Science, University of Science and Technology of China, Hefei 230026, People's Republic of China}

\author[0000-0002-9908-291X]{Yiwei Ni}
\affiliation{School of Astronomy and Space Science, Nanjing University, Nanjing 210023, People's Republic of China}
\affiliation{Key Laboratory of Modern Astronomy and Astrophysics (Nanjing University), Ministry of Education, Nanjing 210023, People's Republic of China}

\begin{abstract}
In this Letter, we perform a detailed analysis of the M5.5-class eruptive flare occurring in active region 12929 on 2022 January 20.
The eruption of a hot channel generates a fast coronal mass ejection (CME) and a dome-shaped extreme-ultraviolet (EUV) wave at speeds of 740$-$860 km s$^{-1}$.
The CME is associated with a type II radio burst, implying that the EUV wave is a fast-mode shock wave.
During the impulsive phase, the flare shows quasi-periodic pulsations (QPPs) in EUV, hard X-ray, and radio wavelengths.
The periods of QPPs range from 18 s to 113 s, indicating that flare energy is released and nonthermal electrons are accelerated intermittently with multiple time scales.
The interaction between the EUV wave and low-lying adjacent coronal loops (ACLs) results in contraction, expansion, and transverse vertical oscillation of ACLs. 
The speed of contraction in 171, 193, and 211 {\AA} is higher than that in 304 {\AA}.
The periods of oscillation are 253 s and 275 s in 304 {\AA} and 171 {\AA}, respectively. A new scenario is proposed to explain the interaction.
The equation that interprets the contraction and oscillation of the overlying coronal loops above a flare core can also interpret the expansion and oscillation of ACLs,
suggesting that the two phenomena are the same in essence.
\end{abstract}


\section{Introduction} \label{intro}
Solar flares and coronal mass ejections (CMEs) are the most spectacular activities in the solar atmosphere, 
during which a huge amount of magnetic free energy is released impetuously \citep{chen11,fle11}. The driver of a CME is mostly a prominence or a flux rope \citep{for00,cx13,vour13,yan18}.
The rapid expansion of a CME during its rising motion leads to the formation of a dome-shaped extreme-ultraviolet (EUV) wave \citep{pat10a,pat10b,ver18}. 
The speeds of EUV waves range from a few hundreds to $\geq$2000 km s$^{-1}$ \citep{shen12,liu18}. 
Furthermore, fast CMEs are capable of driving shock waves propagating in the corona and interplanetary space, 
which are related to type II radio bursts \citep[e.g.,][]{chen02,rou12,zuc18,mor19,down21,lu22,ying22}.

Quasi-periodic pulsations (QPPs) are prevalent in solar and stellar flares \citep[see reviews][and references therein]{vd16,zim21}. 
They are usually detected in hard X-ray (HXR) and radio wavelengths \citep{tan10,hay16}. Sometimes, they are observed in UV and EUV wavelengths \citep{li15,zqm16,tian18}.
QPPs are indicative of intermittent energy release and particle acceleration \citep{cla21}.
The periods of QPPs are between a few seconds to several minutes. Multiple periods in a single flare have been reported \citep{ing09,ning22}.

Kink oscillations are ubiquitous in coronal loops \citep{naka21}. They were first observed by the Transition Region and Coronal Explorer (TRACE) spacecraft in 171 {\AA} images \citep{asch99,naka99}.
The oscillations provide a valuable diagnostics of magnetic field strength and internal Alfv{\'e}n speeds of the oscillating loops \citep{god16}. Transverse kink-mode loop oscillations 
can be excited by flares \citep{wang04,wht12,jain15,zqm20,con22}, coronal jets \citep{dai21}, prominence eruptions \citep{zim15,zqm22}, and EUV waves \citep{shen12}.
\citet{sri13} explored the vertical kink oscillations of a large-scale plasma curtain during the passage of an EUV wave, which is triggered by an X6.9 class flare on 2011 August 9.
\citet{ofm15} carried out 3D MHD modeling of such kind of oscillations, the periods of which are very close to the observed values.
Interestingly, oscillations of global kink mode come out after the implosion of overlying coronal loops during the impulsive phase of flares \citep{go12,sun12,sim13,dud16}.
The abrupt release of magnetic energy leads to a reduction of the loop's support while the inward tension force barely decreases.
As a result, the loop is accelerated by the unbalanced forces and moves downward to reach a new equilibrium \citep{hud00,rus15}.
Before the steep contraction of large-scale loops, there is usually a gradual expansion phase associated with the slow rise of a prominence or a flux rope \citep{liu09,sim13,devi21}.
On the contrary, \citet{chan21} reported loop contraction and subsequent expansion driven by a filament eruption. 
To our knowledge, vertical oscillation during the expanding phase of coronal loops has rarely been investigated.

In this Letter, we report multiwavelength observations of an M5.5-class eruptive flare, which was accompanied with a fast CME on 2022 January 20. 
The flare occurred in NOAA active region (AR) 12929 (N08W76) close to the western limb. 
For the first time, transverse vertical oscillation is discovered in the adjacent coronal loops (ACLs) after sudden contraction and expansion.
The paper is organized as follows. We describe the data analysis in Section~\ref{data}. The results are presented in Section~\ref{res}.
A physical explanation is provided in Section~\ref{dis}. Finally, a brief conclusion is given in Section~\ref{sum}.

\section{Data analysis} \label{data}
The M5.5 flare was observed by the Atmospheric Imaging Assembly \citep[AIA;][]{lem12} on board the Solar Dynamics Observatory (SDO) spacecraft.
AIA took full-disk images in seven EUV (94, 131, 171, 193, 211, 304, and 335 {\AA}) and two UV (1600 and 1700 {\AA}) wavelengths.
The AIA level\_1 data with a time cadence of 12 s and a spatial resolution of 1$\farcs$2 were calibrated using the standard program \texttt{aia\_prep.pro} in the Solar Software (SSW).
Photospheric line-of-sight (LOS) magnetograms of the flare were observed by the Helioseismic and Magnetic Imager \citep[HMI;][]{sch12} on board SDO.
The HMI level\_1 data with a time cadence of 45 s and a spatial resolution of 1$\farcs$2 were calibrated using the standard program \texttt{hmi\_prep.pro}.
The flare was also observed in its decay phase by the H$\alpha$ Imaging Spectrograph (HIS) on board the Chinese H$\alpha$ Solar Explorer \citep[CHASE;][]{li22a}. 
The CHASE/HIS\footnote{https://ssdc.nju.edu.cn} provides H$\alpha$ spectroscopic observations with a pixel spectral resolution of 0.024 {\AA} and a time cadence of one minute \citep{qiu22}.

Irradiance of the flare from a broad band of 1$-$70 {\AA} was directly measured by the EUV SpectroPhotometer (ESP) on board the EUV Variability Experiment \citep[EVE;][]{wood12} of SDO.
Soft X-ray (SXR) fluxes of the flare in 0.5$-$4 {\AA} and 1$-$8 {\AA} were measured by the GOES spacecraft with a cadence of $\sim$2 s.
HXR fluxes at various energy bands were obtained from the Gamma-ray Burst Monitor \citep[GBM;][]{mee09} on board the \textit{Fermi} spacecraft.
Microwave fluxes of the flare were observed by the Nobeyama Radio Polarimeters \citep[NoRP;][]{naka85} with multiple frequencies (1, 2, 3.75, 9.4, 17, 35, 80 GHz).

The related CME was simultaneously observed by the Large Angle and Spectrometric Coronagraph \citep[LASCO;][]{bru95} on board the SOHO spacecraft 
and the COR2 white-light (WL) coronagraph on board the ahead Solar TErrestrial RElations Observatory \citep[STEREO;][]{kai08}. 
The separation angle between the ahead STEREO (hereafter STA) and Earth was $\sim$35$\degr$.
The radio dynamic spectra associated with the flare and CME-driven shock was obtained from the ALMATY ground-based station belonging to the e-Callisto\footnote{http://www.e-callisto.org} network.

\section{Results} \label{res}
\subsection{Eruption of a hot channel} \label{s:hc}
\begin{figure*}
\includegraphics[width=0.90\textwidth,clip=]{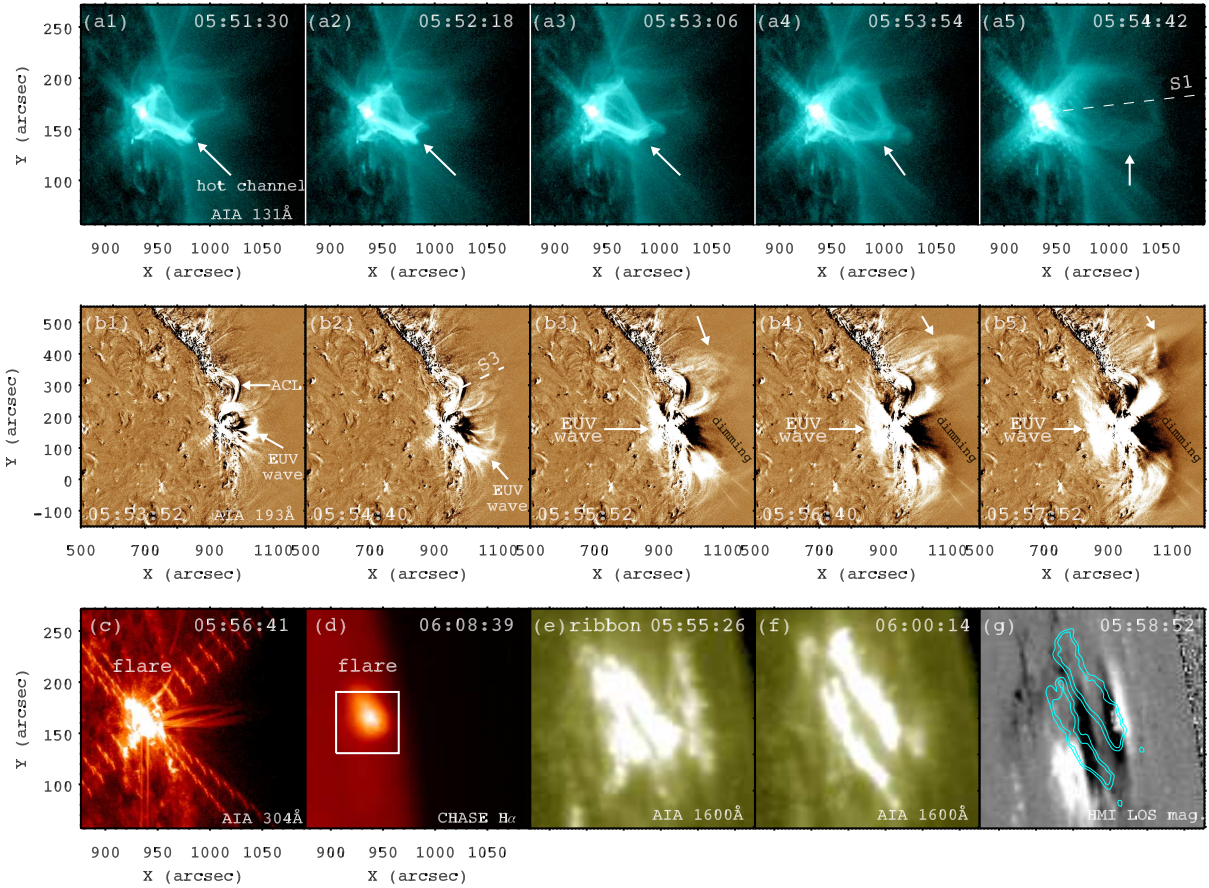}
\centering
\caption{(a1-a5) AIA 131 {\AA} images to illustrate the early eruption of a hot channel. 
In panel (a5), an artificial slice S1 with a length of 180$\arcsec$ is used to investigate the height evolution of the hot channel.
(b1-b5) AIA 193 {\AA} base-difference images to illustrate the formation and propagation of an EUV wave due to the hot channel eruption.
In panel (b1), the horizontal arrow points to the adjacent coronal loops (ACLs). In panel (b2), an artificial slice S3 is used to investigate the evolution of ACLs.
(c-d) The M5.5 flare observed in 304 {\AA} and H$\alpha$. (e-f) Bright flare ribbons observed in 1600 {\AA}. 
(g) Photospheric LOS magnetogram of AR 12929, with the intensity contours of flare ribbons overlaid in cyan lines. The field of view of panels (e-g) is marked with a white box in panel (d). 
An animation showing the eruption of hot channel in AIA 131 {\AA} and the associated EUV wave in AIA 193 {\AA} is available.
It covers a duration of 8 minutes from 05:49:54 UT to 05:57:52 UT on 2022 January 20. The entire movie runs for $\sim$1 s.
(An animation of this figure is available.)}
\label{fig1}
\end{figure*}

In Figure~\ref{fig1}, the top panels show a series of AIA 131 {\AA} images to illustrate the early eruption of a hot channel (see also the online animation \textit{Fig1.mp4}). 
The bright, curved structure rose slowly before 05:51:30 UT. Later on, the rising motion became significant.
The structure is merely evident in 131 {\AA} and 94 {\AA}, but is undistinguishable in other AIA passbands with lower formation temperatures, 
which is in accordance with observational characteristics of hot channels \citep{cx11,wang18}.

\begin{figure}
\includegraphics[width=0.45\textwidth,clip=]{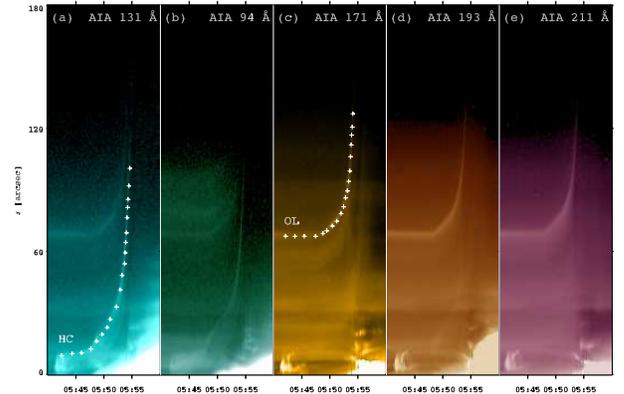}
\centering
\caption{Time-distance plots of S1 in various EUV bands of SDO/AIA.
The trajectory of hot channel (HC) is denoted with ``+'' symbols in panel (a).
The trajectory of the overlying loop (OL) is denoted with ``+'' symbols in panel (c).}
\label{fig2}
\end{figure}

To investigate the height evolution of the hot channel, an artificial slice (S1) along the direction of eruption is selected in Figure~\ref{fig1}(a5).
The time-distance plots of S1 in various EUV bands of AIA are displayed in Figure~\ref{fig2}. 
The trajectory (``+'' symbols) of hot channel in 131 {\AA} and 94 {\AA} is characterized by a slow rise and a fast rise as previously reported.
The overlying loop, $\sim$42.2 Mm above the hot channel, is tardy during the slow rise of hot channel.
It is pushed upward to form the leading front of a CME as the hot channel accelerates \citep{cx13}.
Contrary to the hot channel, the overlying loop is visible in 131 and 94 {\AA}, and is more striking in 171, 193, and 211 {\AA}, suggesting its multithermal nature.

\begin{figure}
\includegraphics[width=0.45\textwidth,clip=]{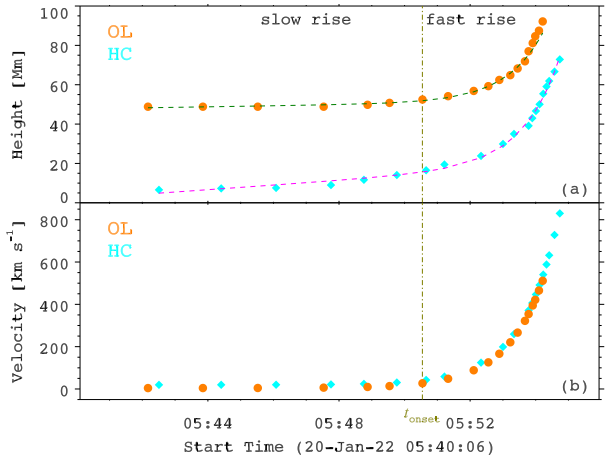}
\centering
\caption{Time evolutions of the heights (upper panel) and velocities (lower panel) of the HC (cyan diamonds) and OL (orange circles).
The olive dash-dotted line indicates the onset time (05:50:30 UT) of HC.}
\label{fig3}
\end{figure}

In Figure~\ref{fig3}(a), time evolutions of the heights of hot channel and overlying loop are plotted with cyan diamonds and orange circles, respectively.
The widely-used function is applied to perform curve fittings:
\begin{equation} \label{eqn-1}
  h(t)=c_0e^{(t-t_0)/\sigma}+c_1(t-t_0)+c_2,
\end{equation}
where $t$ is time, $h(t)$ is height, and $t_0$, $\sigma$, $c_0$, $c_1$, and $c_2$ are free parameters.
The onset time of fast rise is defined as the time when the exponential velocity and the linear velocity are the same, i.e., $t_{\mathrm{onset}}=\sigma\ln(c_1\sigma/c_0)+t_0$.
In Figure~\ref{fig3}(a), the heights of hot channel and overlying loop are fitted with the above function. The results of fittings are superposed with magenta and green dashed lines, respectively.
It is clear that both trajectories can satisfactorily be fitted with this function. The olive dash-dotted line signifies the onset time (05:50:30 UT) of hot channel.
In Figure~\ref{fig3}(b), time evolutions of the velocities of hot channel and overlying loop are displayed with cyan diamonds and orange circles.
Before $t_{\mathrm{onset}}$, the hot channel moves faster than the overlying loop. Afterwards, their velocities are close to each other and the final velocity reaches $\sim$830 km s$^{-1}$.

\subsection{CME and EUV wave} \label{s:cme}
The eruption of a hot channel leads to the formation and propagation of an EUV wave (see also the online animation \textit{Fig1.mp4}). 
Figure~\ref{fig1}(b1-b5) show AIA 193 {\AA} base-difference images, featuring the bright, dome-shaped EUV wave pointed by white arrows.
A dark void or coronal dimming is created behind the eruption as a result of density depletion. 
In Figure~\ref{fig1}(b1), there is a bundle of adjacent coronal loops (ACLs) to the north of flare site. The south footpoints of ACLs are very close to the flare site.
As the EUV wave arrives, compresses, and sweeps the ACLs, the interaction causes contraction, expansion, and oscillation of ACLs, which is described in detail in Section~\ref{s:con}.
With the eruption of a hot channel, the bright flare kernels and ribbons at the chromosphere are simultaneously observed in 
AIA 304 {\AA} and 1600 {\AA} (see the bottom panels of Figure~\ref{fig1}). 
The CHASE/HIS started observing the flare from 06:08:39 UT when H$\alpha$ emission was still significant (see Figure~\ref{fig1}(d)).

\begin{figure*}
\includegraphics[width=0.90\textwidth,clip=]{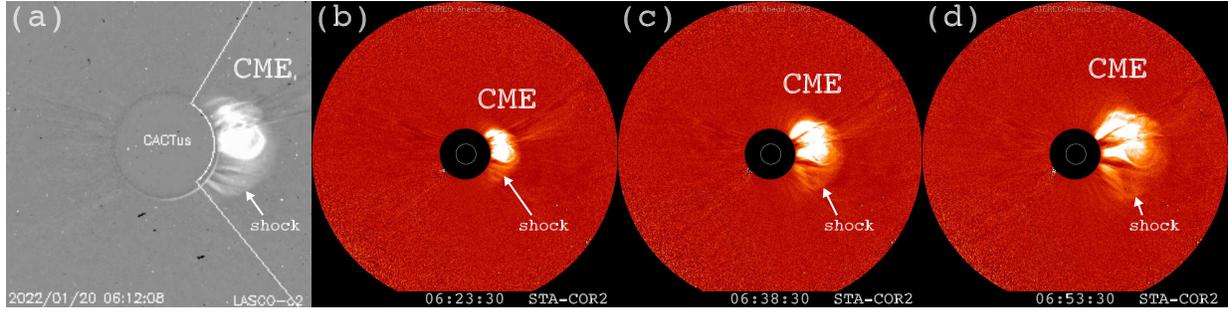}
\centering
\caption{The CME observed by LASCO/C2 at 06:12:08 UT and STA/COR2 during 06:23:30$-$06:53:30 UT. 
The CME-driven shock with lower intensity is pointed by arrows.}
\label{fig4}
\end{figure*}

The successful eruptions of hot channel and overlying loop evolve into a fast and wide CME, which drives a shock wave indicated by the arrows in Figure~\ref{fig4}.
The shape of CME is different from the typical three-part structure \citep{ill85}. 
On the contrary, the resemblance between the CME and hot channel observed in 131 {\AA} implies the flux rope nature of CME.
The apparent speed and angular width of the CME are $\sim$749 km s$^{-1}$ and $\sim$110$\degr$ in the plane of the sky, respectively.
The true (3D) speed of CME is estimated to be $\sim$772 km s$^{-1}$ assuming a radial propagation.

\begin{figure}
\includegraphics[width=0.45\textwidth,clip=]{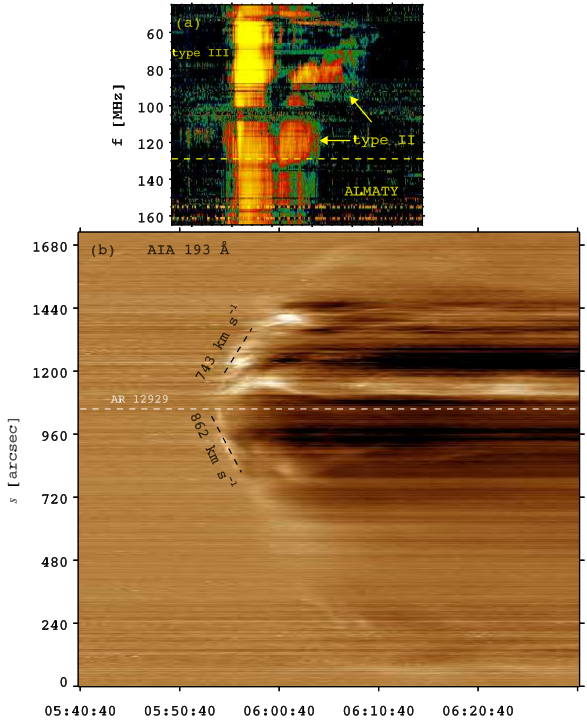}
\centering
\caption{(a) Radio dynamic spectra of the eruptive flare observed by the ALMATY station. 
The type III radio burst during the impulsive phase of flare and the following type II radio burst are pointed by arrows.
The horizontal dashed line is used to extract the radio flux at 127 MHz.
(b) Time-distance plot of S2 in 193 {\AA}. The northward (743 km s$^{-1}$) and southward (862 km s$^{-1}$) speeds of the EUV wave are labeled.}
\label{fig5}
\end{figure}

To calculate the speeds of the EUV wave in the corona, another slice (S2) passing through the wave is selected, which is a quarter of a circle with a radius of 1100$\arcsec$.
The time-distance plot of S2 using the 193 {\AA} base-difference images is displayed in Figure~\ref{fig5}(b). 
The EUV wave propagates northward and southward away from AR 12929 at speeds of 743 km s$^{-1}$ and 862 km s$^{-1}$, respectively.
Figure~\ref{fig5}(a) shows the radio dynamic spectra observed by the ALMATY station. 
A type III radio burst with a fast frequency drift rate during 05:55$-$06:00 UT was coincident with the flare impulsive phase. 
The radio burst was also detected below 16 MHz by the S/WAVES instrument on board STA. A type II radio burst with a slower frequency drift rate followed the type III burst. 
Combining the EUV difference images (Figure~\ref{fig1}(b1-b5)), WL images of CME (Figure~\ref{fig4}), and the type II radio burst, 
it is concluded that the eruption of hot channel evolves into a CME and drives the dome-shaped EUV wave, which is a fast-mode shock in nature.
The speeds of EUV wave in the corona are comparable with the speed of CME.

\subsection{Flare and QPPs} \label{s:qpp}
\begin{figure}
\includegraphics[width=0.45\textwidth,clip=]{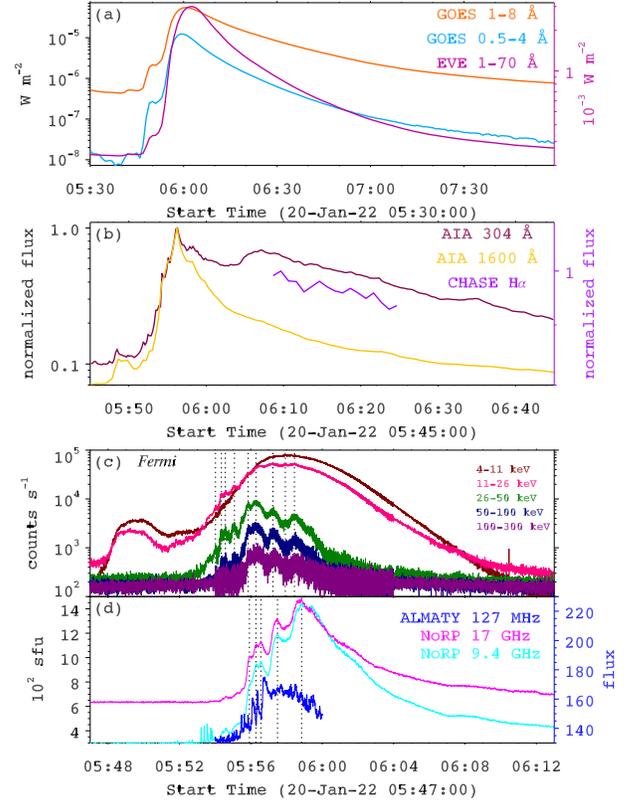}
\centering
\caption{(a) Light curves of the flare in 1$-$8 {\AA}, 0.5$-$4 {\AA}, and 1$-$70 {\AA}.
(b) Normalized fluxes of the flare in 304 {\AA}, 1600 {\AA}, and H$\alpha$.
(c) HXR light curves of the flare at various energy bands (4$-$300 keV).
(d) Radio fluxes of the flare at 9.4 GHz, 17 GHz, and 127 MHz, respectively.}
\label{fig6}
\end{figure}

In Figure~\ref{fig6}, the top panel shows light curves of the M5.5 flare in 1$-$8 {\AA} (orange line), 0.5$-$4 {\AA} (blue line), and 1$-$70 {\AA} (pink line).
The SXR fluxes start to rise at $\sim$05:41 UT and reach the peak at $\sim$06:01 UT, which is followed by a long decay phase until $\sim$08:00 UT.
Therefore, the lifetime of the flare is $\sim$140 minutes. The peak time in 1$-$70 {\AA} is $\sim$06:02:30 UT.
In Figure~\ref{fig6}(b), normalized fluxes of the flare in AIA 304 {\AA} and 1600 {\AA} are drawn with maroon and yellow lines, respectively.
Unlike in SXR, the emissions in these wavelengths have small-amplitude fluctuations during the impulsive phase.
The light curve of the flare in H$\alpha$ line center during 06:08$-$06:24 UT is drawn with a purple line, which has the same descending trend as the 304 {\AA} light curve.

Figure~\ref{fig6}(c) shows HXR light curves of the flare at various energy bands (4$-$300 keV). Variation of the flux at 4$-$11 keV is relatively smooth.
However, the fluctuations become remarkable as energy increases, which is a clear indication of QPPs \citep[e.g.,][]{hay16,cla21}. 
A total of nine peaks are identified (vertical dotted lines) and the peaks are in phase at different energy bands.
In Figure~\ref{fig6}(d), radio fluxes of the flare at 9.4 GHz and 17 GHz are drawn with cyan and magenta lines, respectively.
Similar to the HXR fluxes, the radio fluxes also show QPPs, in which five peaks are identified. As shown in Figure~\ref{fig5}(a), the flare is related to a type III radio burst. 
The fluxes at 127 MHz are extracted from the dynamic spectra (horizontal dashed line) and plotted with a blue line in Figure~\ref{fig6}(d), in which fluctuations are also distinct.

\begin{figure}
\includegraphics[width=0.45\textwidth,clip=]{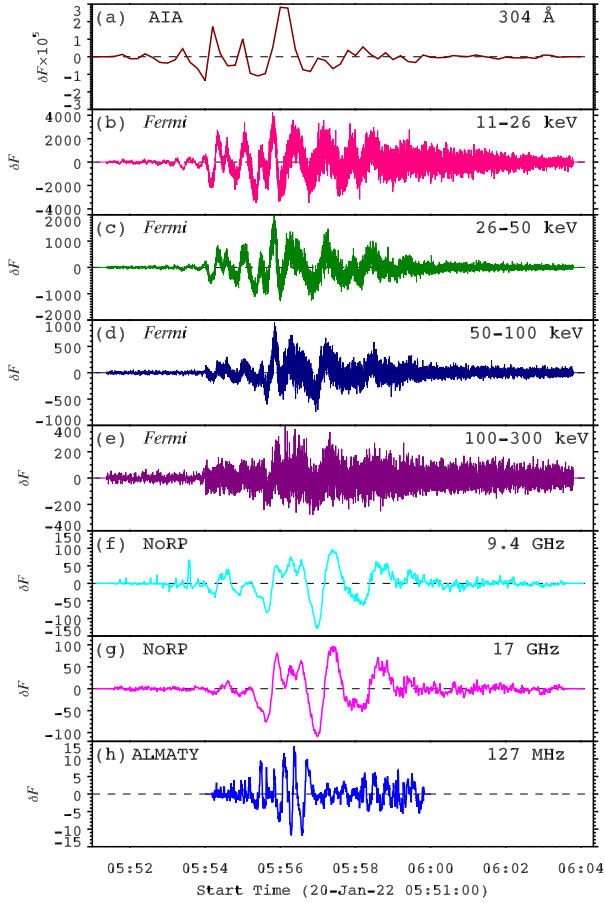}
\centering
\caption{Detrended light curves during the impulsive phase of the flare in EUV, HXR, and radio wavelengths.}
\label{fig7}
\end{figure}

\begin{figure}
\includegraphics[width=0.45\textwidth,clip=]{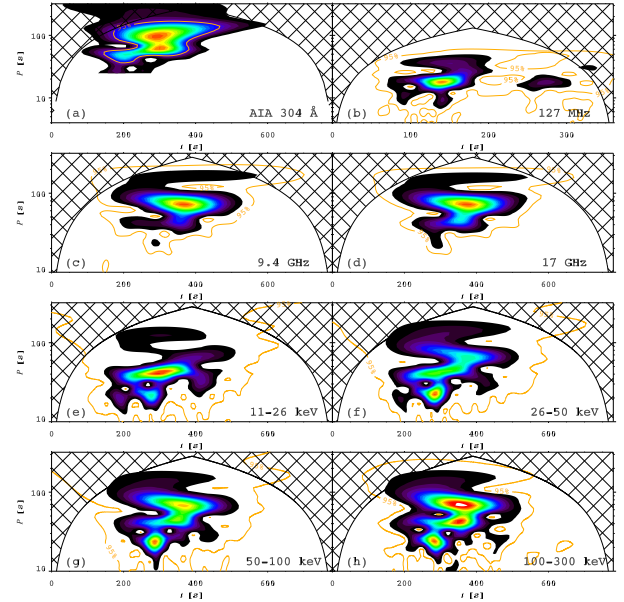}
\centering
\caption{Morlet wavelet transforms of the detrended light curves in Figure~\ref{fig7}.}
\label{fig8}
\end{figure}

To calculate the periods of QPPs in various wavebands, the original light curves are first smoothed to obtain the background intensities, i.e., the slowly varying components.
After subtracting the background intensities from the original light curves, the detrended, fast-varying components are acquired \citep{li15}.
Figure~\ref{fig7} shows the detrended light curves in EUV, HXR, and radio wavelengths. In this way, the peaks identified in Figure~\ref{fig6} are more prominent.
The Morlet wavelet transforms of the detrended light curves are displayed in Figure~\ref{fig8} and the corresponding periods are listed in Table~\ref{tab:qpp}.
In radio wavelengths, there is a unique period of 18 s at 127 MHz and a unique period of 78.6 s at 9.4 and 17 GHz.
In 304 {\AA}, there are two dominant periods, 59.0 s and 113.0 s, the ratio of which is close to 2.
For HXR wavelengths, there is only one period (40.3 s) at 11$-$26 keV. However, there are multiple periods from 23.9 s to 73.8 s at higher energy bands.
QPPs with multiple periods indicate that the flare energy is released and nonthermal electrons are accelerated intermittently with multiple time scales.
The electrons propagating upward generate quasi-periodic emissions during the type III radio burst, while those propagating downward into the chromosphere 
generate quasi-periodic emissions in HXR (11$-$300 keV), microwave (9.4 and 17 GHz), and EUV (304 {\AA}).

\begin{deluxetable}{cccc}
\tablecaption{Periods of QPPs in different wavebands with different cadences. \label{tab:qpp}}
\tablecolumns{4}
\tablenum{1}
\tablewidth{0pt}
\tablehead{
\colhead{Instrument} &
\colhead{Waveband} &
\colhead{Cadence} &
\colhead{Period} \\
\colhead{} &
\colhead{} &
\colhead{(s)} &
\colhead{(s)} 
}
\startdata
SDO/AIA  & 304 {\AA} & 12 & 59.0, 113.0 \\
Fermi/GBM & 11-26 keV & 0.064 & 40.3 \\
Fermi/GBM & 26-50 keV & 0.064 & 23.9, 40.3, 73.8 \\
Fermi/GBM & 50-100 keV & 0.064 & 23.9, 43.9, 73.8 \\
Fermi/GBM & 100-300 keV & 0.064 & 23.9, 40.3, 73.8 \\
NoRP & 9.4 GHz & 1 & 78.6 \\
NoRP & 17 GHz & 1 & 78.6 \\
ALMATY & 127 MHz & 0.25 & 18.0 \\
\enddata
\end{deluxetable}

\subsection{Contraction, expansion, and oscillation of ACLs} \label{s:con}
\begin{figure}
\includegraphics[width=0.40\textwidth,clip=]{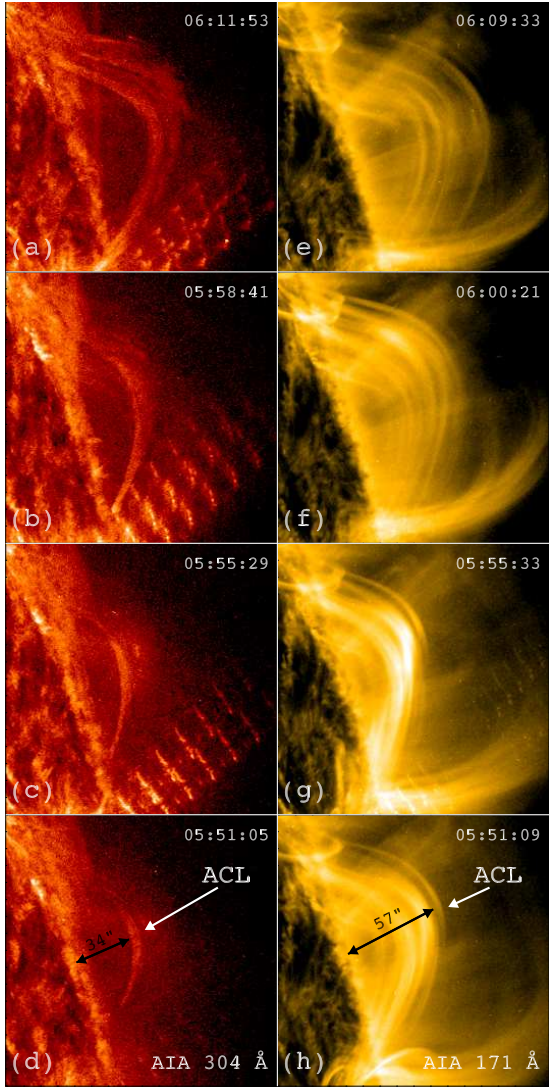}
\centering
\caption{Snapshots of the AIA 304 {\AA} (left panels) and 171 {\AA} (right panels) images during the contraction, expansion, and oscillation phases.
Apparent heights (34$\arcsec$ in 304 {\AA} and 57$\arcsec$ in 171 {\AA}) of ACLs are labeled in the bottom panels.
An animation showing the passage of an EUV wave and the subsequent expansion and transverse vertical oscillation of ACLs in AIA 304 and 171 {\AA} is available.
It covers a duration of 24 minutes from 05:51:05 UT to 06:15:05 UT on 2022 January 20. The entire movie runs for $\sim$4 s.
(An animation of this figure is available.)}
\label{fig9}
\end{figure}

In Figure~\ref{fig1}(b1-b5), the base-difference 193 {\AA} images illustrate the interaction between the fast EUV wave and ACLs.
Figure~\ref{fig9} shows a series of AIA 304 {\AA} and 171 {\AA} image to illustrate the response of ACLs (see also the online animation \textit{Fig9.mp4}).
The bottom panels show the ACLs at the beginning of contraction when the EUV wave arrives and compress the loops. 
As time goes on, the heights of the loops descend to the lowest points around 05:55:30 UT. Meanwhile, the loops brighten due to the strong compression of EUV wave.
The ACLs recover and expand after the passage of EUV wave. 

\begin{figure}
\includegraphics[width=0.45\textwidth,clip=]{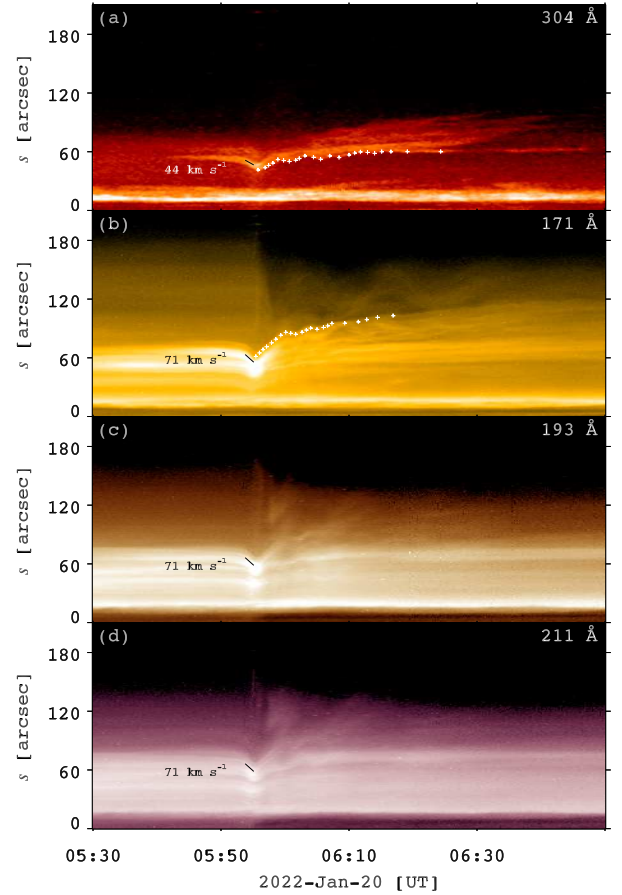}
\centering
\caption{Time-distance plots of S3 in various EUV wavelengths of SDO/AIA. The speeds of contraction (black lines) are labeled. 
The expansion and vertical oscillation of the ACLs in 304 {\AA} and 171 {\AA} are denoted by ``+'' symbols in the top two panels.}
\label{fig10}
\end{figure}

In Figure~\ref{fig1}(b2), a slice (S3) with a length of 210$\arcsec$ is selected to investigate the evolution of ACLs.
The time-distance plots of S3 in various AIA passbands are displayed in Figure~\ref{fig10}. The contraction and expansion of ACLs are obvious and synchronized in all EUV wavelengths.
The speed of contraction in 171, 193, and 211 {\AA} is higher than that in 304 {\AA}. Surprisingly, the ACLs do not come to a halt when recovering to their initial heights. 
Instead, the loops overshoot and oscillate vertically for 4-5 cycles, especially in 304 and 171 {\AA}. 
The heights of new equilibrium are 6$\arcsec$$-$30$\arcsec$ higher than the initial heights before flare.
In 193 and 211 {\AA}, the oscillations are blurred due to the lower contrast with the background corona.

\begin{figure}
\includegraphics[width=0.45\textwidth,clip=]{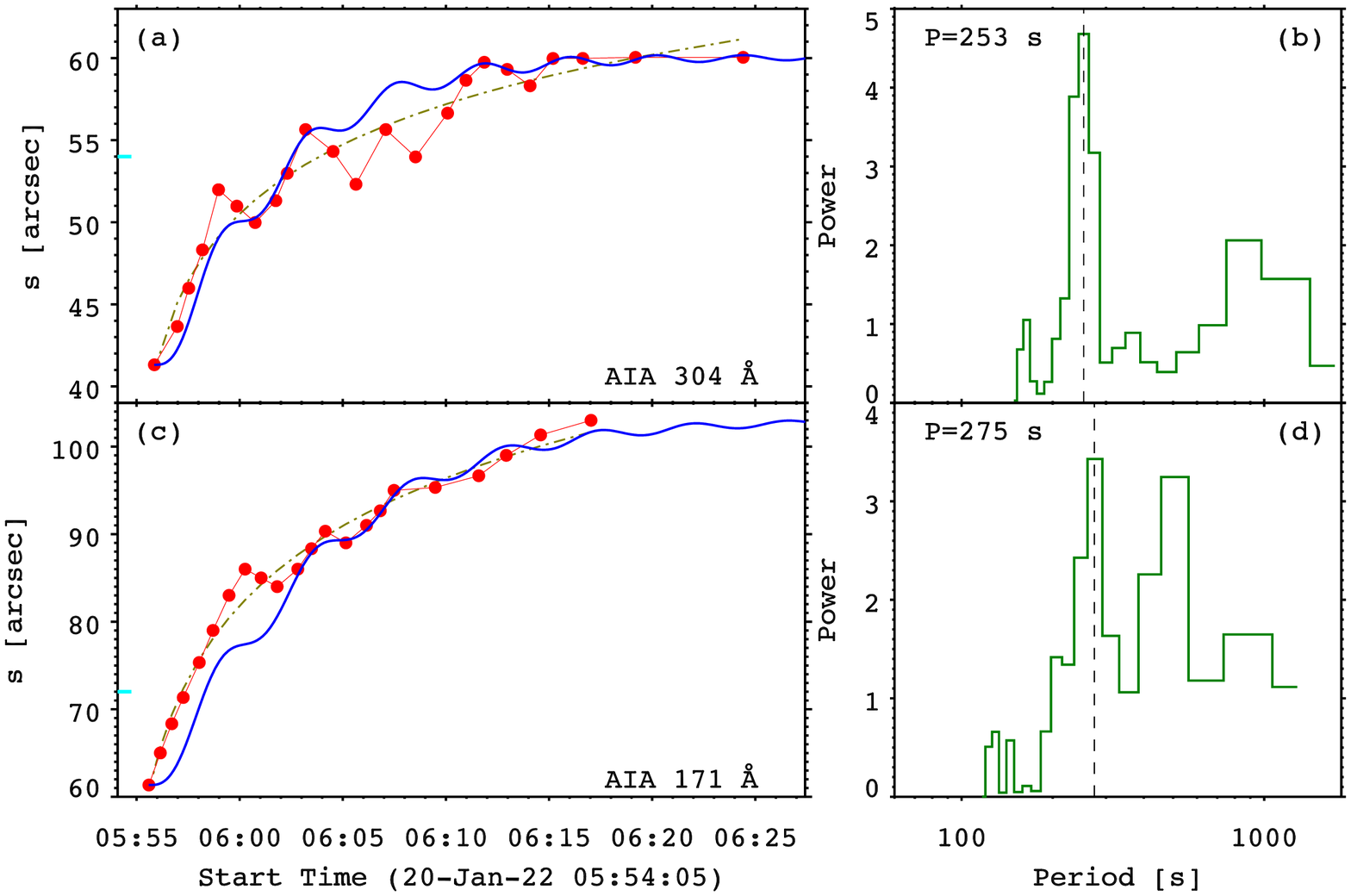}
\centering
\caption{Left panels: Trajectories of ACLs (red circles) during the expansion and oscillation in 304 {\AA} and 171 {\AA}.
The olive dash-dotted lines represent the fitted trends. The short cyan lines mark the initial heights of ACLs before flare.
The blue lines represent the solutions of Equation~(\ref{eqn-3}).
Right panels: Results of fast Fourier transforms of the corresponding detrended trajectories. The periods of vertical oscillations are labeled.}
\label{fig11}
\end{figure}

The height evolutions of ACLs during the expansion and oscillation in 304 and 171 {\AA} are marked with white ``+'' symbols in Figure~\ref{fig10} 
and plotted with red circles in the left panels of Figure~\ref{fig11}. A function is exploited to derive the background trends of loop heights:
\begin{equation} \label{eqn-2}
  h(t)=d_0-\frac{d_1}{(t-t_0)^\alpha},
\end{equation}
where $d_0$ is the final height when $t\gg t_0$, $d_1$ and $\alpha$ are free parameters. The fitted trends are superposed with olive dash-dotted lines in the left panels of Figure~\ref{fig11}.
Fast Fourier transforms (FFT) are performed using the detrended trajectories. The results are plotted in the right panels of Figure~\ref{fig11}.
The periods of vertical oscillations are 253 s and 275 s in 304 {\AA} and 171 {\AA}, respectively.
The difference in period is probably due to the difference in loop length, assuming the same phase speed.
On one hand, the periods of vertical oscillations are much longer than the periods of QPPs (see Table~\ref{tab:qpp}). On the other hand, the flare QPPs occur before oscillations.
Therefore, the QPPs are not modulated by transverse loop oscillations.

\section{Physical explanation of expansion and transverse vertical oscillation} \label{dis}
There are substantial observations and investigations of loop contraction and subsequent vertical oscillation as a result of coronal implosion \citep[e.g.,][]{go12,sun12,sim13,dud16,zqm22}.
To explain the whole process, \citet{rus15} proposed a unified model (``remove-of-support" mechanism). Prior to a flare, inward magnetic tension of the overlying coronal loop is equal to 
the outward magnetic pressure gradient force. During the impulsive phase, the rapid release of magnetic energy reduces the loop's support while the tension force is unaffected.
Hence, the downward net force drives the loop to move toward a new equilibrium, which is accompanied with vertical oscillation.
A modified equation is put forward to describe the contraction and oscillation:
\begin{equation} \label{eqn-3}
  \frac{d^2x}{dt^2}+\omega^2(x-x_0(t))+2\omega\kappa\frac{dx}{dt}=0,
\end{equation}
where $x(t)$ denotes the displacement of the loop, $x_0(t)$ denotes the equilibrium position as a function of time ($t$), $\omega$ denotes the frequency of the undamped oscillation,
and $\kappa$ is the damping ratio. The dimensionless form of Equation~(\ref{eqn-3}) is:
\begin{equation} \label{eqn-4}
  \frac{d^2\tilde{x}}{d\tilde{t}^2}+4\pi^2(\tilde{x}-\tilde{x}_0(\tilde{t}))+4\pi\kappa\frac{d\tilde{x}}{d\tilde{t}}=0,
\end{equation}
where $P=2\pi/\omega$ is the corresponding period of vertical oscillation, $\tilde{t}=t/P$, and $\tilde{x}=x/D$, where $D$ represents the final displacement. 
The following expression of $\tilde{x}_0(\tilde{t})$ is used:
\begin{equation} \label{eqn-5}
  \tilde{x}_0(\tilde{t})=\cases{2, & $\tilde{t}\leq\tilde{t}_c$ \cr
  2-\tanh((\tilde{t}-\tilde{t}_c)/\Delta), & $\tilde{t}>\tilde{t}_c$ \cr}
\end{equation}
where $\Delta=1$. Hence, $\tilde{x}_0(\tilde{t})$ is a decreasing function with $\tilde{t}$ to mimic loop contraction.
The solution of Equation~(\ref{eqn-3}) with $\kappa=0.1$ can well explain the observed quick collapse and oscillation 
when the time scale of change-in-equilibrium is comparable with the oscillation period (see their Fig. 4c).

\begin{figure}
\includegraphics[width=0.45\textwidth,clip=]{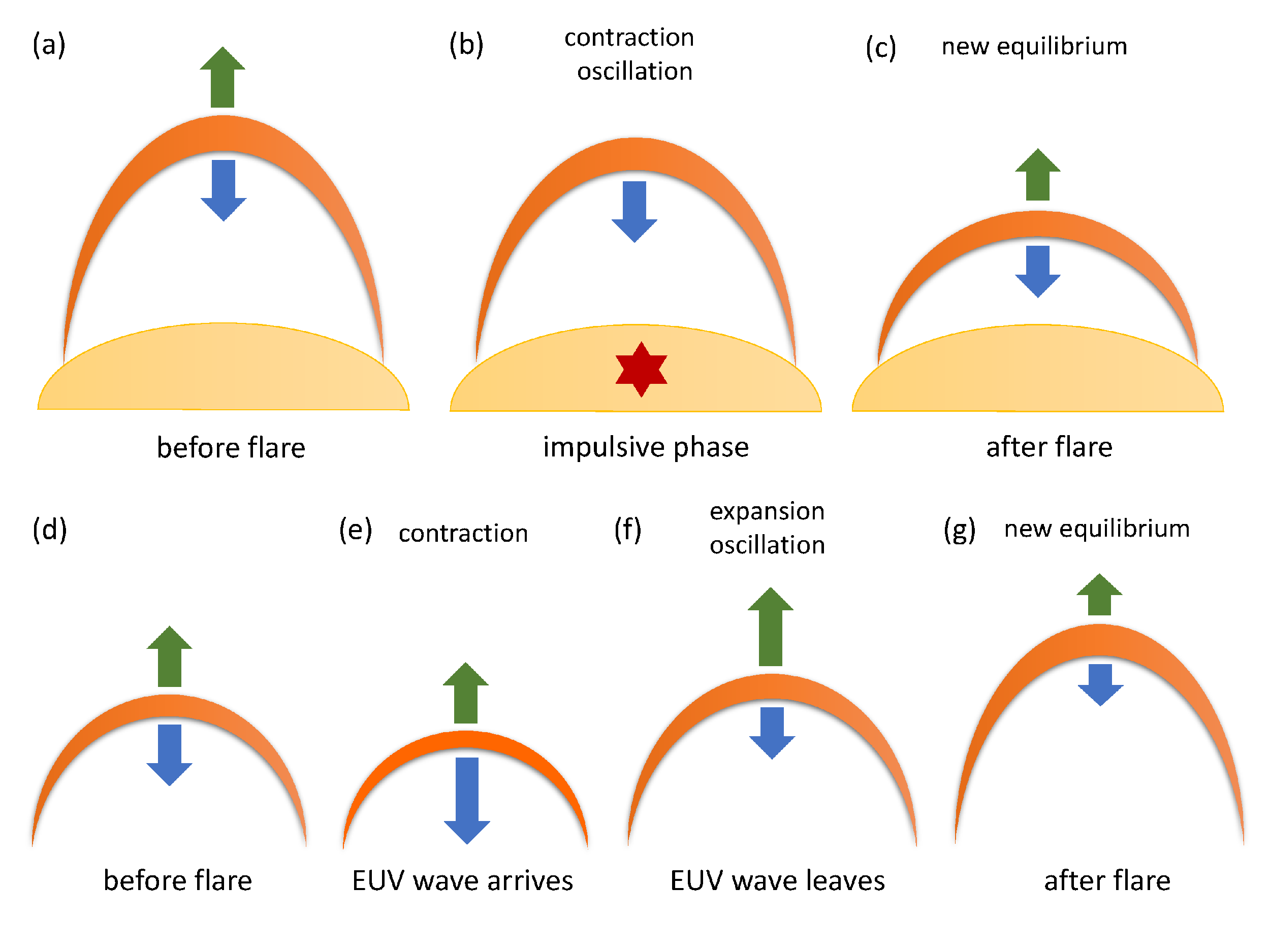}
\centering
\caption{Top panels: A cartoon showing the collapse and oscillation of the overlying coronal loops above flare core.
Bottom panels: A cartoon showing the contraction, expansion, and vertical oscillation of ACLs close to the flare core. See text for detail.}
\label{fig12}
\end{figure}

In our event, the situation is somewhat complicated. 
In Figure~\ref{fig12}, the top panels show a cartoon to illustrate the collapse and oscillation of the overlying coronal loops above the flare core, which is similar to Fig. 3 of \citet{rus15}.
The green and blue arrows represent the outward magnetic pressure gradient force and inward magnetic tension force, respectively. The gravity is neglected in a low-$\beta$ environment.
The bottom panels show a cartoon to illustrate the contraction, expansion, and vertical oscillation of ACLs close to the flare core.
Before flare, the magnetic pressure gradient is balanced by the magnetic tension (panel (d)). 
During the impulsive phase of flare, which is associated with a fast EUV wave, the ACLs are compressed and contract due to a greatly enhanced downward force (panel (e)).
After the passage of EUV wave and ejection of hot channel, the magnetic pressure gradient overtakes the magnetic tension. 
The upward net force pushes ACLs to expand and oscillate vertically in the meanwhile (panel (f)). 
The ACLs reach a new equilibrium eventually, which is higher than the initial position before flare (panel (g)).

To quantitatively describe the expansion and vertical oscillation, we only have to revise $\tilde{x}_0(\tilde{t})$:
\begin{equation} \label{eqn-6}
  \tilde{x}_0(\tilde{t})=\cases{h_0, & $\tilde{t}\leq\tilde{t}_c$ \cr
  h_0+\tanh((\tilde{t}-\tilde{t}_c)/\Delta), & $\tilde{t}>\tilde{t}_c$ \cr}
\end{equation}
where $h_0$ is the height of ACLs before expansion. Hence, $\tilde{x}_0(\tilde{t})$ is an increasing function with $\tilde{t}$ to mimic loop expansion. 
For 304 {\AA}, we set $\kappa=0.05$ and $\Delta=2.0$. Then, Equation~(\ref{eqn-4}) is solved by using the \texttt{ODE45} solver in MATLAB.
Considering $P=253$ s (Figure~\ref{fig11}(b)) and $D=18\farcs$7, the solution of Equation~(\ref{eqn-3}) is derived and superposed in Figure~\ref{fig11}(a) with a blue line.
Likewise, for 171 {\AA}, we set $\kappa=0.05$ and $\Delta=2.5$. Equation~(\ref{eqn-4}) is solved with MATLAB.
Considering $P=275$ s (Figure~\ref{fig11}(d)) and $D=41\farcs$6, the solution of Equation~(\ref{eqn-3}) is derived and superposed in Figure~\ref{fig11}(c) with a blue line.
The periods of oscillations are close to the time scales of change-in-equilibrium (several minutes) during the expansion phase.
The solutions are generally in good agreement with the observed trajectories (red circles) in both wavelengths, including the overall trends and peaks of vertical oscillations,
indicating that Equation~(\ref{eqn-4}) can interpret not only the contraction and oscillation of overlying coronal loops (Figure~\ref{fig12}(a-c)), 
but also the expansion and oscillation of ACLs (Figure~\ref{fig12}(d-g)). There are mainly two reasons for the inconsistence at some data points.
On one hand, the trajectories of ACLs in Figure~\ref{fig10} are obtained manually, which undoubtedly have uncertainties. There might be overlap of loops along the line of sight.
On the other hand, the expression of $\tilde{x}_0(\tilde{t})$ (Equation~(\ref{eqn-6})) is not necessarily the best choice, which should be improved in the future.

Kink oscillations excited by EUV waves are abundant \citep[e.g.,][]{shen12,kum13,sri13}. However, interactions between EUV waves and coronal loops are simple in most cases.
The loops start oscillating once being disturbed by an EUV wave originating from a remote region of eruption. Interaction between the EUV wave and ACLs is more complicated.
When the EUV wave arrives, the ACLs are compressed and undergo rapid contraction, which is followed by expansion and vertical oscillation after the EUV wave sweeps the loops. 
The difference from previous results is probably due to that the ACLs are very close to the flare site and the initial heights of ACLs are relatively low (see bottom panels of Figure~\ref{fig9}).
The velocity and energy of an EUV wave decrease with distance. Consequently, the further an EUV wave propagates, the weaker disturbance a coronal loop has.
It should be emphasized that the HXR loop-top source of a flare also shows contraction and expansion, 
which is explained by fast relaxation of a strongly sheared magnetic field followed by continuing magnetic reconnection \citep{ji06}.

Increasing periods of kink oscillations in expanding coronal loops have been noticed. 
\citet{nis13} investigated the coronal loops in AR 11494 on 2012 May 30. The loops experience small-amplitude decayless oscillations before and well after a C1.0 class flare.
However, the loops show large-amplitude decaying oscillations triggered by the flare and associated CME. 
One of the loop expands gradually with time, during which the period of oscillation increases from $\sim$215 s to $\sim$280 s with a growth rate of 0.43 s/min.

\citet{pas17} performed coronal seismology of a rapidly contracting coronal loop which is oscillating vertically on 2012 March 9 \citep{sim13} 
as well as an expanding loop which is oscillating horizontally on 2012 October 20 \citep{god16}. The loop sways back and forth around a new equilibrium position after a lateral displacement.
The evolution of the period of kink oscillation compared with the background trend is useful in distinguishing between loop motions in the plane of the loop and those perpendicular to it.
In our study, observation of the loop oscillation was available from only one perspective of SDO/AIA. Unfortunately, the flare and ACLs were behind the limb in the FOV of STA/EUVI on that day.
Hence, stereoscopic observation of the loop oscillation was missing to explicitly distinguish whether the contraction is a physical decrease in the lengths of loops or an apparent decrease 
as a result of lateral deflection of the loops. If the initial contraction is caused by lateral deflection, parts of the loops appearing brighter in 171 {\AA} may result from greater overlap along our line of sight,
and the subsequent oscillation would predominantly be horizontal like the event on 2012 October 20.
However, the compression of ACLs by the EUV wave could also lead to increase in plasma density and emission measure in EUV.
On the other hand, the difference images in 171 {\AA} agree with vertical instead of horizontal oscillation according to the result of forward modeling \citep{wang04,wht12}.

Although significant progress has been made in the inference of magnetic field strength and Alfv{\'e}n speeds of oscillating loops using coronal seismology, 
the coronal loops oscillate collectively rather than in isolation. \citet{hind21} derived the skin depth of the near-field response of an oscillating loop, which depends largely on the loop length.
For a typical loop whose axis is semicircular, the skin depth is comparable with the radius of curvature, i.e., the loop height.
In the bottom panels of Figure~\ref{fig9}, the loop heights (34$\arcsec$ and 57$\arcsec$) of ACLs are labeled, which are larger than the apparent separations of the multistranded loops.
Accordingly, the effect of coupling between oscillating loops should be seriously taken into account in the future.

\section{Conclusion} \label{sum}
In this paper, we perform a detailed analysis of the M5.5 class eruptive flare occurring in AR 12929 on 2022 January 20.
The main results are as follows:

\begin{enumerate}
\item{The eruption of a hot channel generates a fast CME and a dome-shaped EUV wave at speeds of 740$-$860 km s$^{-1}$.
The CME is associated with a type II radio burst, implying that the EUV wave is a fast-mode shock wave.}
\item{During the impulsive phase, the flare shows QPPs in EUV (304 {\AA}), HXR (11$-$300 keV), and radio wavelengths (9.4 GHz, 17 GHz, and 127 MHz).
The periods of QPPs range from 18 s to 113 s, indicating that flare energy is released and high-energy electrons are accelerated intermittently with multiple time scales.}
\item{Interaction between the EUV wave and low-lying ACLs results in contraction, expansion, and transverse vertical oscillation of ACLs.
The speed of contraction in 171, 193, and 211 {\AA} is higher than that in 304 {\AA}.
The periods of oscillations are 253 s and 275 s in 304 and 171 {\AA}, respectively. A new scenario is proposed to explain the interaction.
The equation that interprets the contraction and oscillation of the overlying coronal loops above a flare core can also interpret the expansion and oscillation of ACLs,
suggesting that the two phenomena are the same in essence.}
\item{Similar interaction between EUV waves and nearby filaments is highly expected \citep{li22b}. 
Numerical simulations are urgently desirable to reproduce this process and justify our scenario \citep{ofm15}.}
\end{enumerate}

\begin{acknowledgments}
The authors appreciate the referee for valuable and constructive suggestions.
We thank Drs. J. S. Zhao and L. Chen in Purple Mountain Observatory for their kind help.
The e-Callisto data are courtesy of the Institute for Data Science FHNW Brugg/Windisch, Switzerland.
SDO is a mission of NASA\rq{}s Living With a Star Program. AIA and HMI data are courtesy of the NASA/SDO science teams.
The CHASE mission is supported by China National Space Administration (CNSA).
This work is supported by the National Key R\&D Program of China 2021YFA1600500 (2021YFA1600502) and NSFC grants (No. 11790302, 11790300, 11973092, 12073081).
\end{acknowledgments}

%\bibliography{sample631}{}
%\bibliographystyle{aasjournal}

\end{document}